\documentstyle[epsfig,11pt]{article}

\title{A Study of the Unidentified Gamma-Ray Source 3EG
J1828+0142}
\author{Jorge A. Combi, Gustavo E. Romero, Diego F. Torres,\\
{\small Instituto Argentino de Radioastronom\'{\i}a, C.C.5, 1894
Villa Elisa, Buenos Aires, Argentina}
 \\ \& Brian Punsly \\ {\small  4014 Emerald Street, No. 116,
Torrance, CA 90503, USA}}

\begin{document}

\maketitle

\begin{abstract}
We present a study of the radio environment of the gamma-ray
source 3EG J1828+0142. This source presents a very high
variability index in its gamma emission and a steep high-energy
spectral index $\Gamma\sim-2.8$. Large-scale radio maps at
different frequencies show the presence of a shell-type
non-thermal structure when the diffuse emission from the Galactic
disk is eliminated. At small scales, VLA radio images reveal the
existence of several weak point sources within the 95 \% gamma-ray
confidence location contour. Most of these sources present
non-thermal spectral indices, but it is not clear whether they are
extragalactic or not. We discuss the possibility of a scenario
where the large structure is a SNR and one of the compact radio
sources is an isolated black hole also produced in the original
supernova explosion. The black hole could be responsible for the
variable gamma-ray emission according to Punsly's (1998) model.
INTEGRAL observations with IBIS imager could detect the inverse
Compton and blueshifted pair annihilation radiation from the
relativistic electron-positron jets of the hole. Some estimates
are presented in this regard.

\end{abstract}

\section{Introduction}

The best estimated position of the gamma ray source 3EG J1828+0142
is at $(l\;b)\approx(31.9^{\circ},\;5.8^{\circ})$, according to
the Third EGRET catalog (Hartman et al. 1999). There are not known
potential galactic counterparts like hot massive stars, supernova
remnants (SNR) or young pulsars within the positional error box
(Romero et al. 1999). The source is very variable over timescales
of months (Torres et al. 2000). No strong radio source that could
be identified with a blazar is found within the 95 \% confidence
contour of the gamma-ray emission. The high-energy spectral index
$\Gamma=-2.76\pm0.39$ ($F\propto E^{\Gamma}$) is steeper than what
is expected for most galactic gamma-ray sources. All these
features make of 3EG J1828+0142 a very puzzling object.

In this paper we present both large and small scale radio maps of
the region where this high-energy source is located. These maps
could provide some clues about the origin and nature of 3EG
J1828+0142.

\section{Large-scale radio maps}

We have studied the surroundings of 3EG J1828+0142 using radio
data from the large-scale surveys by Haslam et al. (1981) and
Reich \& Reich (1986). Small-scale VLA observations from the NVSS
Sky Survey by Condon et al. (1998) were used for a complementary
study of the emission within the 95 \% confidence location
contour. We have applied a well-proven Gaussian filtering
technique to the radio images (e.g. Combi et al. 1999) in order to
remove the background contamination from the large-scale
non-thermal disk emission.

In Figure 1, upper panel, we show a map of the radio field around
3EG J1828+0142 at 1.4 GHz. The probability location contours of
the gamma-ray source are superposed to the radio image. A large,
shell-type structure can be clearly seen. The outer boundary has
been encircled in the figure. It is a weak source (the integrated
flux density is $18.2\pm2.1$ Jy at 1.4 GHz) with a low surface
brightness that very much resembles a typical SNR. The
identification is confirmed by the non-thermal spectral index
found for the radio emission: $\alpha=-0.72\pm0.18$
($S_{\nu}\propto\nu^{\alpha}$). The gamma-ray source is located on
the boundary of this new SNR.

\begin{figure}
\epsfig{figure=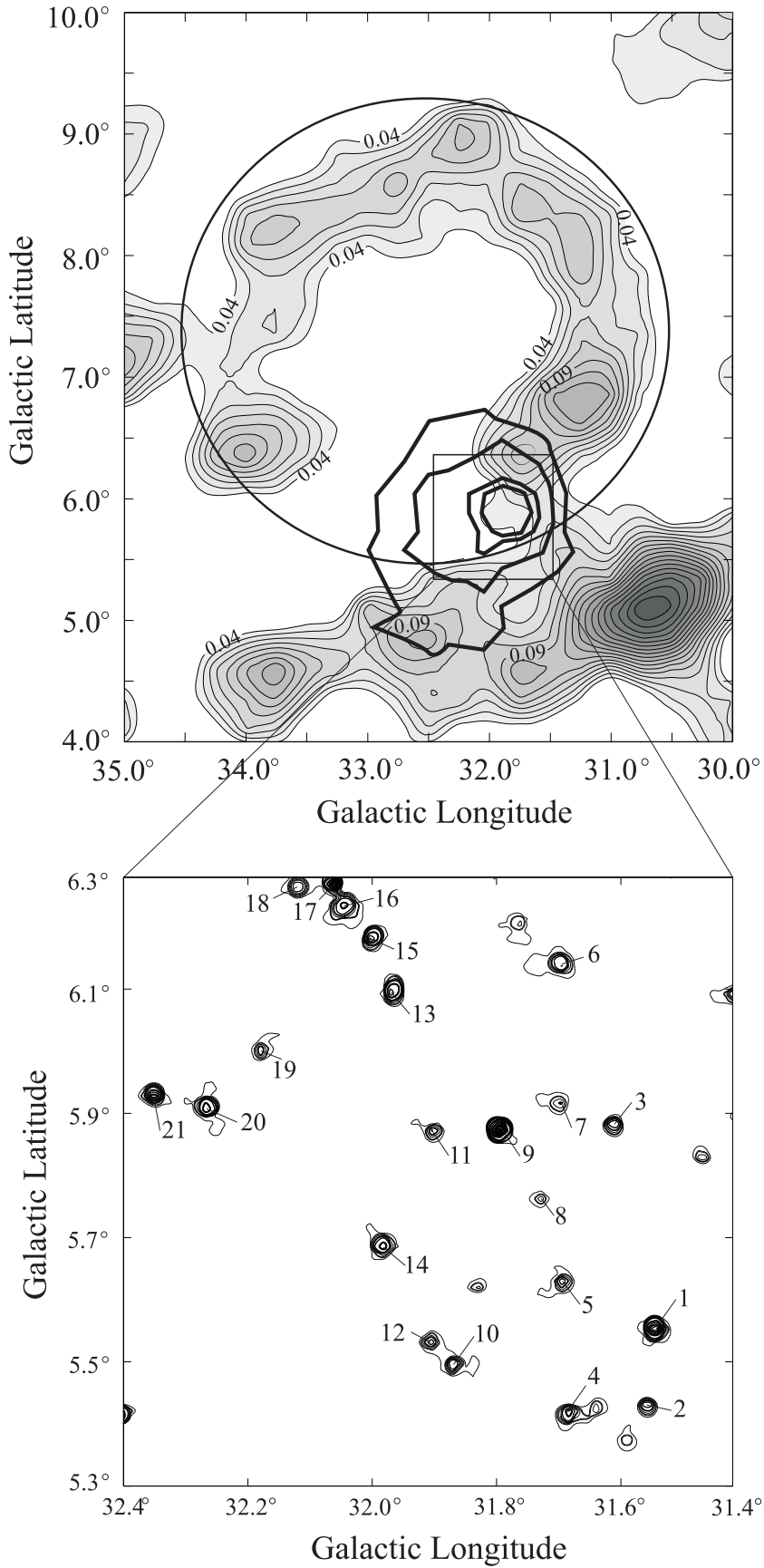} \caption{{\bf Upper panel}: 1.4-GHz
continuum map of the region around 3EG J1828+0142. A Gaussian
filtering beam of $90´\times 90´$ has been applied to remove the
diffuse background emission. Radio contours are shown in steps of
0.01 K starting at 0.04 K. The noise level is 0.02 K. Probability
confidence contours for the location of the gamma-ray source have
been superposed. {\bf Lower panel}: Point-like radio sources at
the same frequency, detected by the VLA within the 95 \%
probability confidence contour of the gamma-ray
source.\label{fig1}}
\end{figure}

The fact that the gamma-ray emission is highly variable rules out
the possibility that the source be the result of the interaction
of the SNR with a cloudy medium as it has been suggested for other
unidentified gamma-ray sources (e.g. Combi et al. 1998). A pulsar
origin is also discarded by the variability levels and the steep
spectrum. The gamma-ray emitting object, however, could be a
compact remnant left by the SN explosion if it is a black hole
instead of a pulsar. An isolated black hole would not undergo
steady accretion from a stellar companion and consequently it
should not be a strong X-ray source. Moreover, if the whole is
charged and a strong magnetic field is supported by an orbiting
opposite-charged ring or disk, as suggested by Punsly (1998), the
system could produce relativistic leptonic jets, which can emit
inverse Compton (IC) and annihilation gamma-rays (see Romero et
al., these proceedings, for additional details of the model). The
jets would also generate synchrotron radio emission, and hence the
source should have a weak non-thermal counterpart.

\section{Point-like radio sources in the field}

In order to find the potential radio counterpart of the gamma-ray
source we have used VLA images of the region within the EGRET 95
\% confidence contour. The lower panel of Figure 1 shows the
point-like radio sources visible at 1.4 GHz with flux densities
above 5 mJy. In Table 1 we list the characteristics of these
sources. Most of them have no entry in any current catalog. We
have computed spectral indices when observations at other
frequencies were available. For instance, source No. 9 corresponds
to TXS B1825+016, with a flux density of 278 mJy at 365 Mhz
(Douglas et al. 1996). Source No. 11 has been detected at 3900 MHz
with a flux of 61 mJy (Torres et al. 2000) and source No. 15 is at
the position of PNM J1827+0201, which presents a flux of $\sim49$
mJy at 4.85 GHz (Griffith \& Wright 1993). For those sources not
detected at 5 GHz in Condon et al.'s (1994) survey we have
estimated upper limits for the spectral index. Many sources have
non-thermal spectra and can be considered as potential
radio-counterparts of the gamma-ray source, if the latter is
produced by a magnetized black hole.

\begin{table}
\caption{Point radio sources inside the 95 \% $\gamma$-ray
confidence contour. (Flux density measured at 1.4
GHz)}\vspace{1em}
\renewcommand{\arraystretch}{1.2}
\begin{tabular}[h]{lllr}
\hline Source No & Coordinates & Flux & $\alpha$
\\
          &  (l,b)      &   (mJy)        &                     \\
\hline 1   & (31.53,+5.56) &   51.2  & $<$ -1.8        \\ 2   &
(31.55,+5.44) &   14.6  & $<$ -0.84        \\ 3   & (31.6,+5.89) &
15.6  & $<$ -0.89       \\ 4   & (31.67,+5.42) &   32.9  & $<$
-1.4        \\ 5   & (31.69,+5.64) &   12.8  & $<$ -0.74
\\ 6   & (31.7,+6.14)  &   35.9  & $<$ -1.5        \\ 7   &
(31.7,+5.92)  &   10.6  & $<$ -0.59         \\ 8   & (31.72,+5.77)
&    7.30  & $<$ 0.3         \\ 9   & (31.78,+5.88) &   79.7  &
-0.9      \\ 10  & (31.86,+5.5)  &   22.0  & $<$ -1.23
\\ 11  & (31.89,+5.88) &   13.8  &  1.34     \\ 12
& (31.89,+5.54) &   28.1  &  0.34     \\ 13  & (31.96,+6.1) & 46.4
& $<$ -1.75
\\ 14  & (31.98,+5.69) &   25.1  & $<$ -1.2  \\ 15  &
(31.99,+6.19) &   43.5  &  0.10     \\ 16  & (32.04,+6.24) & 69.1
&  0.48     \\ 17  & (32.07,+6.29) & 84.2 & $<$ -2.2
\\ 18  & (32.12,+6.27) &   18.7  &  $<$ -1.0
\\ 19  & (32.17,+6.0)  &   12.6  & $<$ -0.72
\\ 20  & (32.27,+5.91) &   29.4  &  $<$ -1.3
\\ 21  & (32.35,+5.93) &   41.2  &  $<$ -1.65       \\ \hline \\
\end{tabular}
\label{tab:table}
\end{table}

\section{Discussion}

The large angular size of the new SNR suggests that it is a nearby
source. Applying the $\Sigma-D$ relationship derived by
Allakhverdiyev et al. (1988) for shell-type remnants of low
surface brightness, we estimate a distance of about 1 kpc. The
radius of the remnant, then, would be $\sim 30$ pc and its age
$\sim 4\times 10^4$ yr, assuming standard values for the particle
density of the ISM and the original energy release (Spitzer 1998).
Since the gamma-ray source is located on the boundary of the
remnant, the hypothetical black hole should have a birth velocity
of $\sim 700$ km s$ ^{-1}$. Although high, such a velocity is not
unreasonable in the light of the recent estimates of Lyne \&
Lorimer (1994) for the birth velocities of radio pulsars.

\begin{figure} \vspace{-0.5cm}
\includegraphics[width=0.8\linewidth]{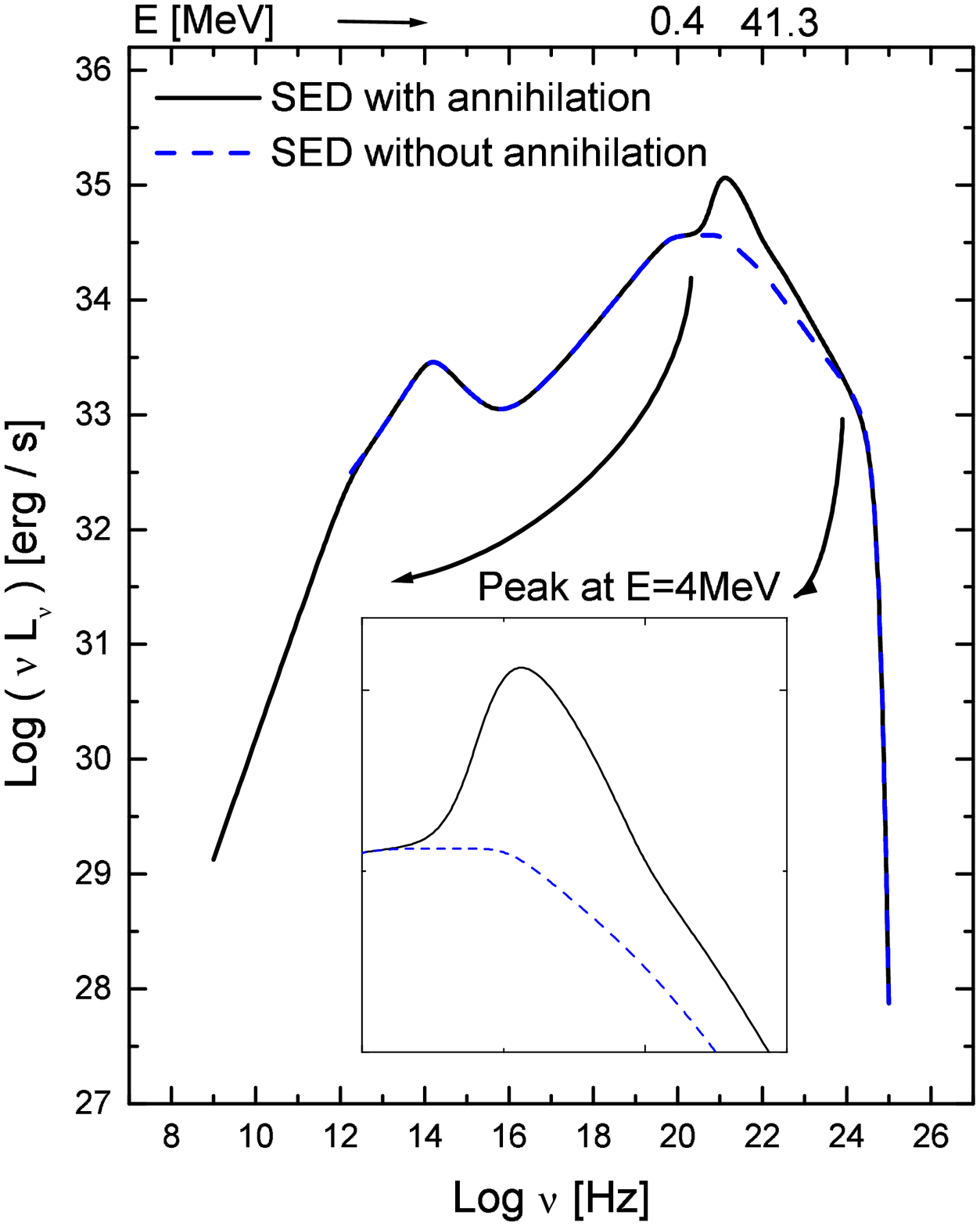}
\caption{Spectral energy distribution for a
magnetized black hole of 4 $M_{\odot}$ with a field of $10^{11}$
G.\label{fig2}}
\end{figure}

The spectral energy distribution of a magnetized black hole in
isolation (see Figure 2) has a broad peak at MeV energies, where
the blueshifted annihilation radiation exceeds the synchrotron
self-Compton emission. Adopting a set of standard parameters for
the hole ($M=4$ $M_{\odot}$, $B=10^{11}$ G), the peak luminosity
results $\sim 1.8 \times 10^{35}$ erg s$ ^{-1}$ at $\sim 6$ MeV
(Punsly et al. 2000). At a distance of 1 kpc this implies a flux
on Earth of $\sim 1.5 \times 10^{-4}$ ph cm$^{-2}$ s$^{-1}$. The
source, consequently, should be detected by the Imager IBIS of the
INTEGRAL mission. Frequent observations could provide more
information on the variability of the emission. Spectral
decomposition in the range 15 KeV -- 10 MeV should reveal the
existence of a luminosity peak at a few MeV. The exact position of
the peak can be used to determine the Doppler factor of the beamed
radiation. The improved source location will also allow to
restrict the number of possible radio counterparts given in Table
1.

\section{Conclusions}

We report the discovery of a new and large SNR, till now masked by
the diffuse background radio emission of the Galaxy. The variable
gamma-ray source 3EG J1828+0142 is located on the outer boundary
of this remnant. We suggest that this source could be a magnetized
black hole produced by the original supernova explosion. The hole
would have a transverse birth velocity of $\sim700$ km s$^{-1}$
and an age of $\sim4\times10^{4}$ yr. Gamma-rays are expected from
electron-positron annihilations and synchrotron self-Compton
losses in the jets generated by the hole. A detailed model (Punsly
et al. 2000) predicts a peak luminosity in the spectral energy
distribution of $\sim 1.8 \times 10^{35}$ erg s$ ^{-1}$ at $\sim
6$ MeV. Imager IBIS of the INTEGRAL mission could test this
hypothesis on the nature of 3EG J1828+0142 through:
\begin{itemize}
\item Detecting the predicted peak of the spectral energy
distribution at MeV energies.
\item Providing a better source location that enables to
identify the expected weak radio counterpart.
\item Measuring with high confidence the level of variability.
\end{itemize}
If the identification with a magnetized black hole is supported by
the observations, then IBIS data can be used for:
\begin{itemize}
\item Estimating the actual value of the Doppler factor of the jet flow (from the
blueshifted annihilation peak).
\item Estimating the particle density in the jet (from the
integrated annihilation line luminosity).
\item Clarifying the mechanism that produce the gamma-ray variability.
\end{itemize}

At present time, it is not known whether charged and rotating
(Kerr-Newman) black holes can exist in isolation. If stable
configurations of black hole plus magnetosphere as those proposed
by Punsly (1998) actually occur in the universe, then some
variable gamma-ray sources like 3EG J1828+0142 could be their
high-energy manifestations. The INTEGRAL mission, with the fine
imaging capability and spectral sensitivity of IBIS instrument,
could be a fundamental key to unveil the existence of such
mysterious objects.

\subsection*{ACKNOWLEDGEMENTS}

This work has been supported by CONICET, ANPCT (PICT No. 03-04881)
and Fundaci\'on Antorchas. G.E.R. is very grateful to the
organizers for a travel grant that made possible his participation
in the Workshop.

\end{document}